\documentclass[a4paper,aps,prl,twocolumn,longbibliography,english,superscriptaddress]{revtex4-2}

\usepackage{amssymb}
\usepackage{graphicx}
\usepackage{dcolumn}
\usepackage{bm}
\usepackage{multirow}
\usepackage{amsmath}
\usepackage{subfigure}
\usepackage{float}
\usepackage{color}
\DeclareUnicodeCharacter{202C}{ }
\usepackage[colorlinks=true,	linkcolor=blue,urlcolor=blue,anchorcolor=blue,citecolor=blue,bookmarksnumbered]{hyperref}

\begin{document}

\title{Direction switchable single-photon emitter using a Rydberg polariton}

\author{Changcheng Li}
\thanks{These authors contributed equally to this work.}
\affiliation{State Key Laboratory of Quantum Optics Technologies and Devices, Institute of Laser Spectroscopy, Shanxi University, Taiyuan 030006, China}

\author{Xiao-Feng Shi}
\thanks{These authors contributed equally to this work.}
\affiliation{Center for Theoretical Physics and School of Physics and Optoelectronic Engineering, Hainan University, Haikou 570228, China}

\author{Yuechun Jiao}
\email{ycjiao@sxu.edu.cn}
\affiliation{State Key Laboratory of Quantum Optics Technologies and Devices, Institute of Laser Spectroscopy, Shanxi University, Taiyuan 030006, China}
\affiliation{Collaborative Innovation Center of Extreme Optics, Shanxi University, Taiyuan 030006, China}

\author{Jiuheng Yang}
\affiliation{State Key Laboratory of Quantum Optics Technologies and Devices, Institute of Laser Spectroscopy, Shanxi University, Taiyuan 030006, China}

\author{Jingxu Bai}
\affiliation{State Key Laboratory of Quantum Optics Technologies and Devices, Institute of Laser Spectroscopy, Shanxi University, Taiyuan 030006, China}
\affiliation{Collaborative Innovation Center of Extreme Optics, Shanxi University, Taiyuan 030006, China}

\author{C. Stuart Adams}
\affiliation{Joint Quantum Centre (JQC) Durham-Newcastle, Department of Physics, Durham University, Durham DH1 3LE, United Kingdom}

\author{Suotang Jia}%
\affiliation{State Key Laboratory of Quantum Optics Technologies and Devices, Institute of Laser Spectroscopy, Shanxi University, Taiyuan 030006, China}
\affiliation{Collaborative Innovation Center of Extreme Optics, Shanxi University, Taiyuan 030006, China}

\author{Jianming Zhao}%
\email{zhaojm@sxu.edu.cn}
\affiliation{State Key Laboratory of Quantum Optics Technologies and Devices, Institute of Laser Spectroscopy, Shanxi University, Taiyuan 030006, China}
\affiliation{Collaborative Innovation Center of Extreme Optics, Shanxi University, Taiyuan 030006, China}

\date{\today}
\begin{abstract}
All-optical redirection or routing of single photons is essential for quantum networks. Although studied in various systems both in theory and experiment, the redirection of single photons with many output ports, compatible with large-scale photonic circuits, still needs to be explored. Here, we demonstrate a direction switchable single-photon emitter using a Rydberg polariton.
The Rydberg component of the stored photon is changed using a stimulated Raman transition with a specific intermediate state. By adjusting the direction of the retrieval laser, we can redirect the emitted photon into a rich variety of alternative modes. Building upon this scheme, we propose a 
quantum routing of single photons with \textit{N} output channels and unity routing efficiency. In addition, the protocol reduces the effect of motional dephasing increasing the photon lifetime to $>10~\mu$s ($>20$ times photon processing time), enabling functional quantum devices based on Rydberg polaritons.

\end{abstract}
\maketitle
Photons are ideal carriers of quantum
information due to their weak interaction with the environment. Consequently, photons are excellent candidates for transferring and manipulating quantum information~\cite{monroe2002, northup2014}. The realization of scalable quantum networks~\cite{kimble2008} requires a quantum router. 
Thus, a single-photon routing scheme, which can achieve dynamic reconfiguration and interface with quantum memories, has received considerable attention in the past decade. For this purpose, numerous theoretical and experimental researches on photon quantum routing have been demonstrated in various systems, such as coupled-resonator waveguides~\cite {PhysRevLett.111.103604, PhysRevA.89.013805,PhysRevA.99.033827}, whispering gallery resonators~\cite{PhysRevLett.102.083601, PhysRevX.3.031013, shomroni2014b}, waveguide-emitter systems~\cite{ yan2014single,PhysRevA.92.063836, zhang2022controllable}, and superconducting qubits~\cite{PhysRevLett.107.073601, PhysRevLett.120.063603, PhysRevApplied.15.014049}. However, the realization of quantum routers with a large number of output channels still needs to be explored.

Rydberg atomic systems have emerged as a dynamic playground for quantum optics \cite{Firstenberg_2016}. An advantage of Rydberg systems is that the strong interaction between Rydberg excitations can be mapped into an interaction between photons and hence an optical non-linearity at the single photon level. Examples of Rydberg single-photon devices include single-photon sources~\cite{dudin2012Strongly,peyronel2012Quantumb, Maxwell2013, ornelas-huerta2020Ondemand, padron-brito2021Probing}, transistors~\cite{baur2014SinglePhotonb,tiarks2014SinglePhoton,gorniaczyk2014SinglePhoton}, photonic quantum gate~\cite{tiarks2019Photon, shi2022Highfidelityb}, photonic entanglement filter~\cite{ye2023Photonica}, contactless photon-photon interactions~\cite{busche2017Contactlessb} and collectively encoded qubits~\cite{spong2021collectively}. Recently, single-photon routers in interacting Rydberg gas have been proposed in theory, such as the quantum reflection of single photons~\cite{murray2017a,ou2022}, chiral quantum router~\cite{palaiodimopoulos2024}, and Rydberg-mediated cavity QED~\cite{du2021}. 
However, a practical demonstration of a Rydberg-mediated single-photon router and how this could be used in the optical quantum network has not been fully explored.

\begin{figure*}[htbp]
    \centering  \includegraphics[width=\linewidth]{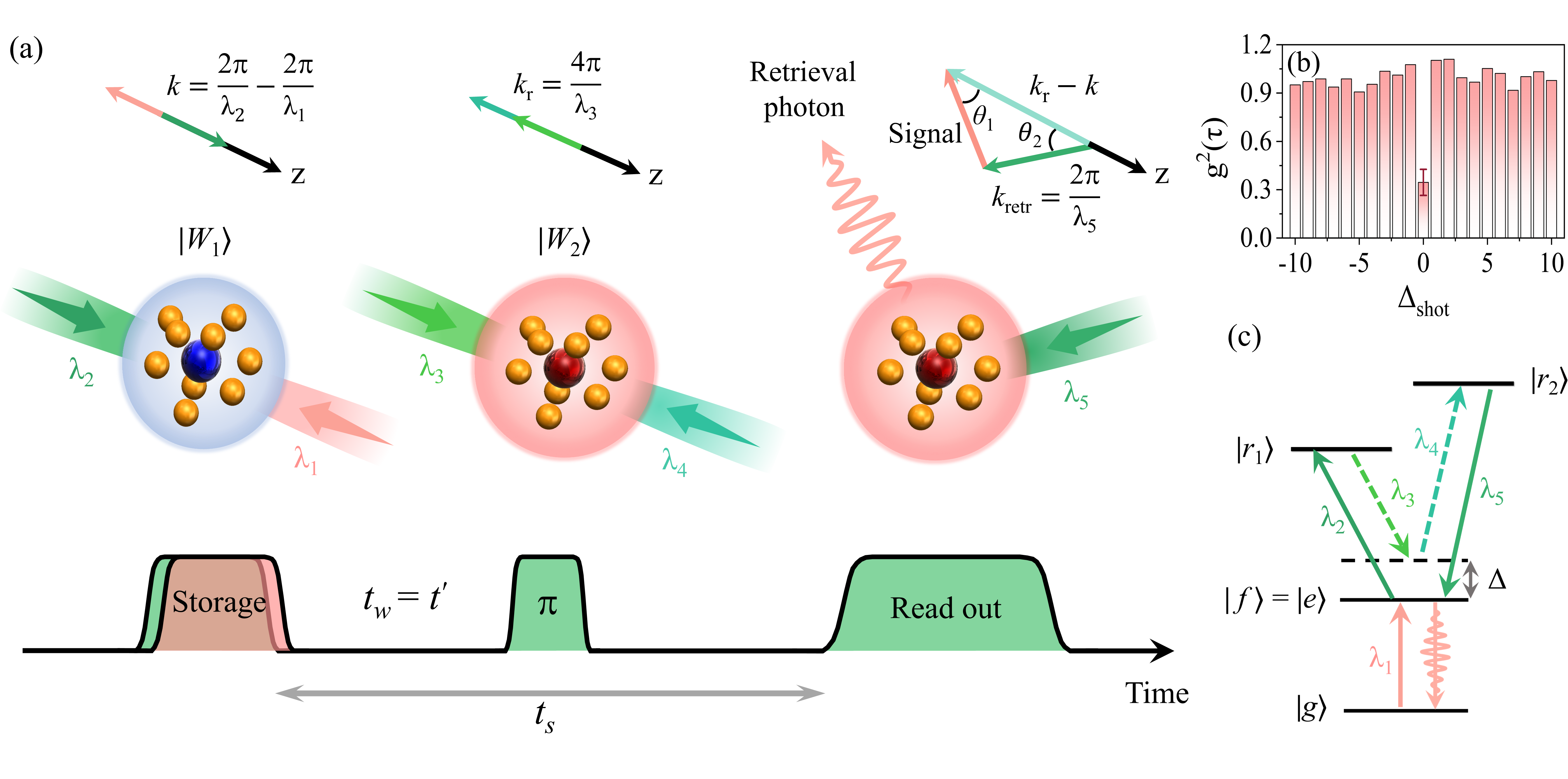}   \caption{(a) Realization of the $\pi$-redirection protocol. (b) Measurement of the second-order correlation function of the retrieval photons. The measured $g^{(2)}(0)$ = 0.34 ± 0.08 for the retrieval signal at Rydberg state $n$ = 70 while $t_s$ = 7~$\mu$s . (c) Relevant energy levels. The signal photon is stored in a Rydberg state $\vert r_1\rangle$ in the form of $\lvert W_1\rangle$ using two-photon Rydberg EIT, where $\lambda_1$ (852~nm) and $\lambda_2$ (509~nm) resonantly couple the two-photon transition $|g\rangle \rightarrow |r_1\rangle$ via the intermediate state $|e\rangle$. Two Raman lasers ($\lambda_{3(4)}$) convert the $|r_1\rangle$ to $|r_2\rangle$ via the intermediate state $|f\rangle=|e\rangle$ with a large detuning $\Delta$, converting to $\lvert W_2\rangle$. The stored photon is read out by the retrieval laser $\lambda_5$ that resonantly couples the $|r_2\rangle \rightarrow |e\rangle$ and detected by a single-photon detector. It appears possible to read out the signal only if the wavevectors of the retrieval laser, readout signal, and the $\lvert W_2\rangle$ form a closed triangle. The angles $\theta_1$ and $\theta_2$ are determined by vectors of $k$, $k_\text{r}$, the retrieval laser $k_\text{retr}$, and the readout signal. The experiments demonstrate a proof-of-principle example with $\theta_1 =0^{\circ}$ and $\theta_2 =0^{\circ}$ (case 7 in Table~\ref{table3}) where the retrieval laser travels exactly opposite to the coupling laser during loading. 
    }
    \label{Fig.1}
\end{figure*}

In this work, we propose a protocol to develop a Rydberg polariton based single-photon router that simultaneously exhibits suppression of motional dephasing. The scheme 
builds on previous work~\cite{PhysRevLett.134.053604}, 
however, instead of two $\pi$ pulses, we only use a single $\pi$ pulse to convert the Rydberg polariton $\vert W_1\rangle$ to another Rydberg polariton $\vert W_2\rangle$ via a specific
intermediate state. This $\vert W_2\rangle$ polariton can no longer be read out due to a mismatch between the wavevectors~\cite{Fleischhauer2002}. However, if we redirect the read-out retrieval laser, we can restore the phase matching condition and achieve efficient retrieval of the stored photon. 
Importantly, this protocol means that the retrieved signal field is coupled into a mode with a different $k$-vector. 
By varying the direction of the retrieval fields, we achieve a direction-switchable single-photon emitter, a key component for a quantum router. The details of the retrieval laser direction and the specific atomic states involved are detailed in Methods. Here, we implement a proof-of-principle configuration by retrieving the signal with the direction of the retrieval laser that is exactly opposite to that of the loading stage. Remarkably, our results show that for long storage times over $10~\mu$s, the efficiency can be more than twice than that of earlier work on the cancellation of motional dephasing~\cite{PhysRevLett.134.053604}.
This opens up exciting perspectives for functional quantum devices based on Rydberg polaritons in single photon switching and routing.

{\it Theoretical model --}
To illustrate how the seemingly impossible redirected readout becomes possible, consider the collective Rydberg polariton $|W_1\rangle$ defined as, 
\begin{eqnarray}
|W_1\rangle = \frac{1}{\sqrt{N}}\sum_{j=1}^{N}e^{ikz_j(0)} |gg\cdots r_1^{(j)}ggg\rangle,\label{Wstate}
\end{eqnarray}
where $k$ is the effective wavevector of the excitation lasers, $z_j(0)$ is the initial position of atom $j$, $\lvert g\rangle$ denotes an atom in the ground state and $\lvert r_1^{(j)}\rangle$ denotes that atom-$j$ is in an excited state $\lvert r_1\rangle$. After using a $\pi$ Raman lasers of wavevector $k_\text{r}$ to convert the Rydberg state to $\lvert r_2\rangle$ via an intermediate state $|f\rangle$, 
Eq.~(\ref{Wstate}) becomes

\begin{eqnarray}
|W_2\rangle = \frac{1}{\sqrt{N}}\sum_{j=1}^{N}e^{i(k-k_{\text{r}})z_j(0) - ik_{\text{r}}v_j(t'+\pi/2\Omega_{\text{r}})  } |gg\cdots r_2^{(j)}ggg\rangle,\nonumber\\\label{Wstate2}
\end{eqnarray}
where $t'$ is the time between the completion of the loading of the signal and the start of the $\pi$ pulse. There are two thermal motion induced random parameters for each atom, the initial position $z_j(0)$ and the velocity $v_j$. According to the picture of collective radiation required for a coherent retrieval of the signal photon, the phase term in Eq.~(\ref{Wstate2}) should be equal to $k'z_j(t_s)$, where $k'=k_\text{r}-k$ is the wavevector of $|W_2\rangle$ and $z_j(t_s)=z_j(0)+v_jt_s$ is the position of the atom $z_j(t_s)$ at the onset time $t_s$ of the retrieval of the signal. The puzzle here is how to convert the phase term in Eq.~(\ref{Wstate2}) which has two random parameters $z_j(0)$ and $v_j$ for each atom into another phase parameter $k'z_j(t_s)$ which only has one random parameter $z_j(t_s)$. 

We find, surprisingly, that one can use redirected pathways for the retrieval laser field to provide a solution to this puzzle, thus also leading to the redirection of the read-out signal field. A cartoon of the protocol is shown in Fig.~\ref{Fig.1}(a).
We label the new method as the $\pi$-redirection protocol that both redirect the read-out signal and cancel the motional dephasing of the polariton.
After writing the signal photon into the polariton mode $|W_1\rangle$, we apply a $\pi$ pulse at time 
\begin{eqnarray}
t'=t_s\left(1- \frac{k}{k_{\text{\tiny{r}}}}\right) - \frac{\pi}{2\Omega_{\text{\tiny{r}}}}.\label{protocol3-1}
\end{eqnarray}

During the free evolution time between photon storage and the $\pi$ pulse, nothing changes to the internal state of the atoms, but they move in real space plus a small acceleration due to gravity. This motion leads to a mismatch between the desired phase in the wavefunction and the actual phase. To achieve coherent radiation at a later time $t_s$ for atomic positions $z_j(t_s)$, the phase term in Eq.~(\ref{Wstate2}) should be $k'z_j(t_s)$. However, this phase is {\it a priori} unknown. Even so, this can be possible only if
\begin{eqnarray}
 k-k_{\text{\tiny{r}}}&=&-k_{\text{\tiny{r}}} \frac{ \frac{\pi}{2\Omega_{\text{\tiny{r}}}} +t' }{×t_s}.
 \label{eq4}
\end{eqnarray}
Once the condition in Eq.~(\ref{eq4}) is satisfied, Eq.~(\ref{Wstate2}) can be written as 
\begin{eqnarray}
\lvert W_2\rangle &=&\frac{-i}{\sqrt{N}} \sum_{j=1}^{N}e^{ik' (z_j(0) + v_j t_s)} |gg\cdots r_2^{(j)}\cdots ggg\rangle. \nonumber\\
\label{protocol3-6}
\end{eqnarray}

This means that as long as we can retrieve the signal with a new wavevector $k'$ at the retrieval time $t_s$, no motional dephasing occurs. We note that compared to the theory in Ref.~\cite{PhysRevLett.134.053604}, half the amount of the Raman laser field is used here, which reduces the noise level of Raman lasers, and hence the coherence time should be enhanced. 

\begin{table}
  \centering
  \begin{tabular}{c|c|c|c|c|c|c|c}
    \hline
    \hline   Case & $k_{\text{\tiny{r}}}/{2k}$ & $\lambda_1/$nm  & $\lambda_2/$nm &  $\lvert f\rangle$ & input & Signal & $(\theta_1,\theta_2)$ \\ \hline

1&	1.05& 894.59 & 495.08 & $ 7P_{3/2} $ &    \multirow{2}{*}{
  \begin{minipage}[c]{0.095\columnwidth}
    \centering
    \includegraphics[width=\linewidth]{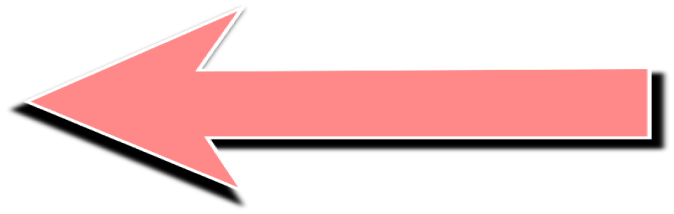}
  \end{minipage}
    }    &  \multirow{2}{*}{
  \begin{minipage}[c]{0.09\columnwidth}
    \centering
    \includegraphics[width=\linewidth]{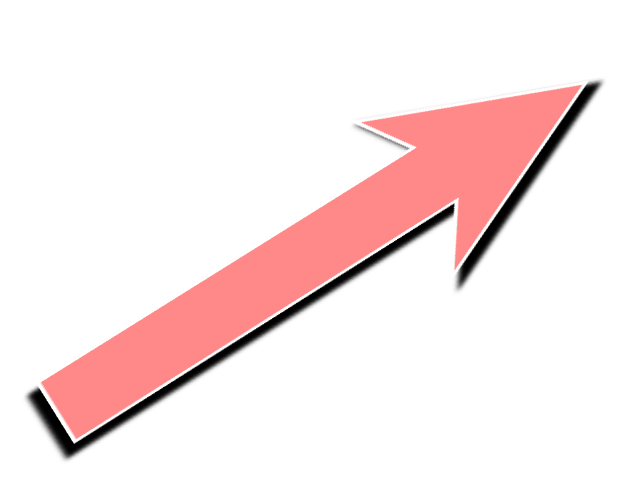}
  \end{minipage}
    }    & (2.58,0.30)  \\
2&	1.07& 894.59 & 495.08 & $ 7P_{1/2} $ &      &       & (2.47,0.35)	 \\\hline
3&	1.19& 852.35 & 509.04 & $ 7P_{3/2} $ &    \multirow{2}{*}{
  \begin{minipage}[c]{0.095\columnwidth}
    \centering
    \includegraphics[width=\linewidth]{input.png}
  \end{minipage}
    }    &    \multirow{2}{*}{
  \begin{minipage}[c]{0.051\columnwidth}
    \centering
    \includegraphics[width=\linewidth]{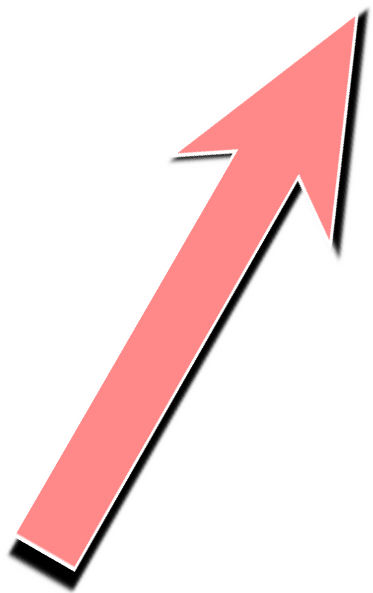}
  \end{minipage}
    }   & (2.09,0.54) 	\\ 
4&	1.21& 852.35 & 509.04 & $ 7P_{1/2} $ &       &    & (2.04,0.56) 	 \\\hline
5&	2.18& 894.59 & 495.08 & $ 6P_{3/2} $ &    \begin{minipage}[b]{0.095\columnwidth}
		\centering
		\raisebox{-.15\height}{\includegraphics[width=\linewidth]{input.png}}
	\end{minipage}    &    \begin{minipage}[b]{0.095\columnwidth}
		\centering
		\raisebox{-.15\height}{\includegraphics[width=\linewidth]{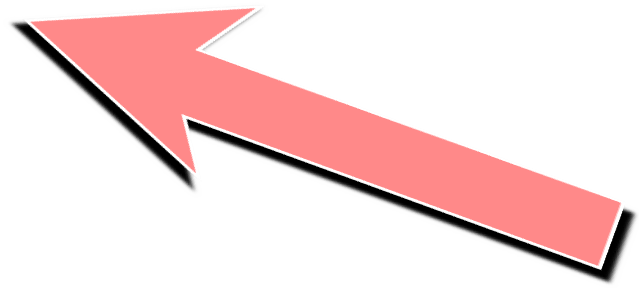}}
	\end{minipage}   & (0.36,0.20) \\\hline
6&	2.24& 894.59 & 495.08 & $ 6P_{1/2} $ &   \multirow{2}{*}{
  \begin{minipage}[c]{0.095\columnwidth}
    \centering
    \includegraphics[width=\linewidth]{input.png}
  \end{minipage}
    }     &    \multirow{2}{*}{
  \begin{minipage}[c]{0.095\columnwidth}
    \centering
    \includegraphics[width=\linewidth]{input.png}
  \end{minipage}
    }   & (0,0)	\\
7&	2.48& 852.35 & 509.04 & $ 6P_{3/2} $ &        &       & (0,0) \\  \hline
\hline
  \end{tabular}
  \caption{ \label{table3} Level configurations and the read-out direction of a signal for the $\pi$-redirection protocol. The arrows present the direction of the input photon and the read-out signal. Here, we use $\lambda_5 \approx \lambda_2$.}
  \end{table}
  
Importantly, by employing a new direction for the
retrieval laser, we can achieve a direction switchable single-photon emitter with the read-out direction depending on $(k_\text{r}-k)\vec{z}-\vec{k}_\text{retr}$, i.e., it is possible to read out the signal only if the wavevectors of the retrieval laser $\vec{k}_\text{retr}$, readout signal, and the $\lvert W_2\rangle$ form a closed triangle, shown in Fig.~\ref{Fig.1}(a).
Table~\ref{table3} displays 7 possible cases of readout directions with the read-out signal frequency same with that of the incoming photon, where the wavelength of the retrieval laser is almost the same as the coupling laser during the loading stage, i.e., $\lambda_5 \approx \lambda_2$. 
We show that case 5 leads to the read-out signal in a direction slightly deviated from its incoming direction, and cases 1-4 result in the read-out of the retrieved signal in a direction nearly opposite to that of the signal photon loaded in the gas. In these cases, if we rotate the retrieval laser around the wavevectors of the $\lvert W_2\rangle$, the read-out signal photon will be rotated to match the relation in the triangle, thus achieving the single-photon quantum routing with generic \textit{N} output channels. For cases 6-7, the retrieved signal will come out in the direction as that of in the loading with the retrieval laser field opposite to that during the loading. Note that the angles $\theta_1$ and $\theta_2$ depend on both the intermediate state $|f\rangle$ coupled by the Raman lasers ($\lambda_{3(4)}$) and the read-out transition coupled by the retrieval laser $\lambda_5$, see details in Methods. The Raman lasers ($\lambda_{3(4)}$) counter-propagate with each other while the $\lambda_3$ field co-propagates with the coupling laser $\lambda_2$, shown in Fig.~\ref{Fig.1}(a). Moreover, if we use a different retrieval laser, i.e., $\lambda_5 \neq \lambda_2$, we can achieve both redirection read-out and conversion of photonic frequency simultaneously, see details in Methods. 

\begin{figure}[htbp]
    \centering    \includegraphics[width=\linewidth]{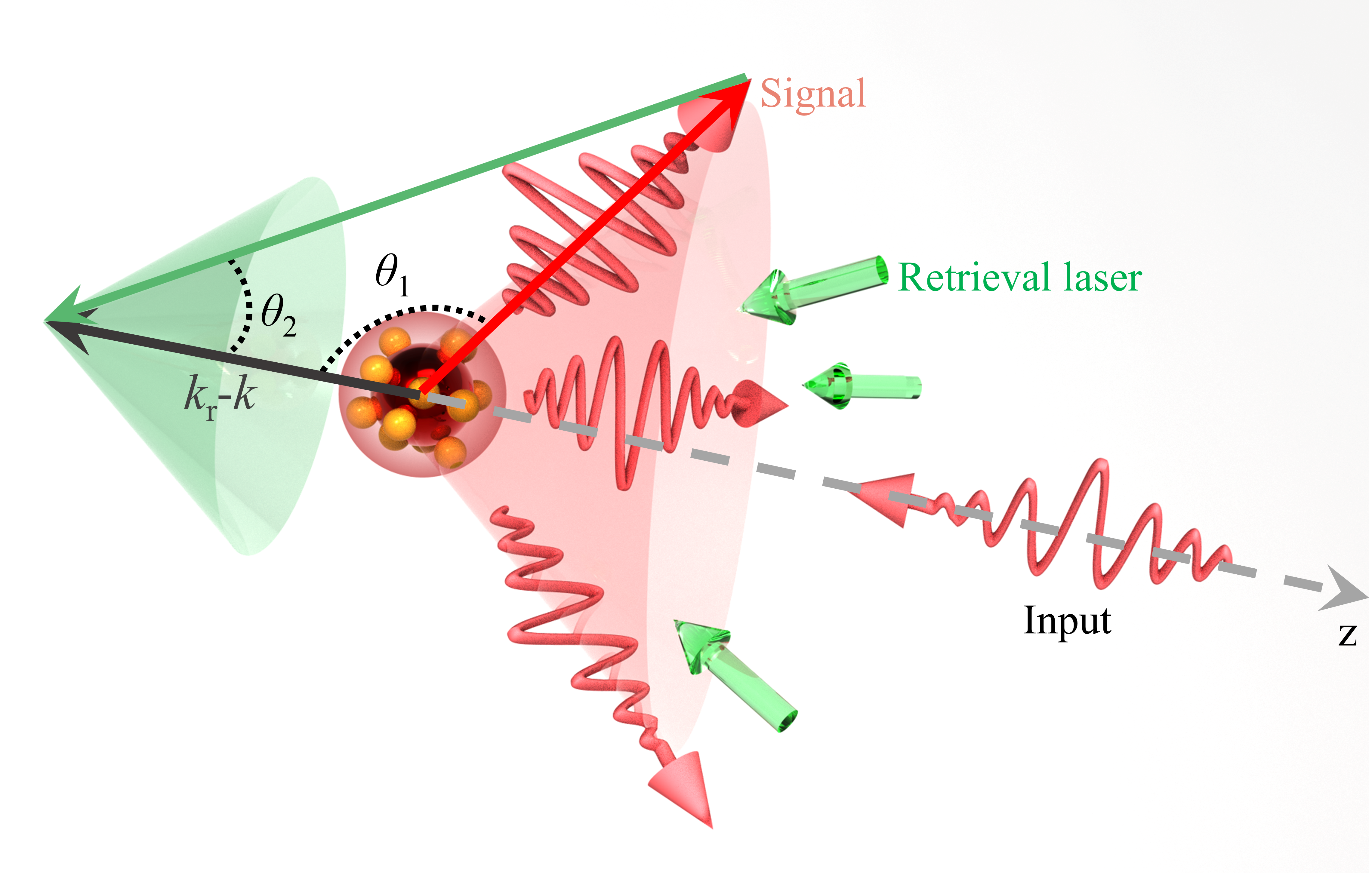}
\caption{Proposed single-photon router with \textit{N} output channels at specific case 4 with $\theta_1 = 2.04$~rad and $\theta_2 = 0.56$~rad. The rotation of the retrieval laser around the wavevectors ($k_\text{r}-k$) of the $\lvert W_2\rangle$  leads to the corresponding routing of the signal photon. }
    \label{Fig.4}
\end{figure}

Based on the new protocol of the redirection photon, we propose a Rydberg-mediated single-photon quantum router with case 4 in Table~\ref{table3}, where we send the retrieval laser in a direction with $\theta_2 = 0.56$~rad, leading the signal photon come out with $\theta_1 = 2.04$~rad. By rotating the retrieval laser around the wavevectors of the $\lvert W_2\rangle$, a single-photon quantum routing with generic \textit{N} output channels is achieved, shown in Fig.~\ref{Fig.4}. Remarkably, the proposed single-photon quantum router has a unity routing efficiency through the redirection retrieval laser.

Limited by the Raman laser at hand, below, we experimentally demonstrate the new protocol using case 7 in Table~\ref{table3}, where $\lambda_2, \lambda_3, \lambda_4$ and $\lambda_5$ are roughly same, to rigorously verify the feasibility of the protocol that can realize the direction switchable readout of the single photon by redirected retrieval laser, meanwhile enhances the coherence time of our Rydberg photon memory as expected.

{\it Experimental realization --} 
The experimental protocol and relevant energy levels are shown in Figs.~\ref{Fig.1}(a) and (c), respectively. A quasi-one dimensional ensemble of Cs atoms was loaded into a 1064~nm optical dipole trap with the $1/e^2$  waist $\emph{w}_1 = 4.3~\mu$m. A 650~ns signal pulse ($\lambda_1$) with a mean photon number of about 1.0 is focused into the ensemble with a $1/e^2$ waist of $\emph{w}_p = 2.5~\mu$m. This resonantly drives the  $|g\rangle=|6S_{1/2}, F=4, m_{F}=4\rangle\rightarrow|e\rangle=|6P_{3/2}, F'=5, m_{F'}=5\rangle$ transition. A strong coupling laser ($\lambda_2$) with Rabi frequency $\Omega_2 = 2\pi \times 7$~MHz and a beam waist $\emph{w}_c = 8.5~\rm{\mu}$m drives the transition $|e\rangle \rightarrow |r_1\rangle=|65S_{1/2}\rangle$, converting the signal photon into a Rydberg polariton, $|W_1\rangle$. The signal and coupling lasers have opposite circular polarizations ($\sigma^+$ and $\sigma^-$) and counter-propagate through the atomic ensemble. The signal photon is stored by ramping down the intensity of the coupling laser before the signal laser is switched off. Due to the Rydberg blockade, only one polariton can be stored~\cite{li2016Quantuma,Adams2020}, shown by the $g^2(\tau)$ measurement of Fig.~\ref{Fig.1}(b). During the storage of the signal, we implement the $\pi$-redirection protocol shown in Fig.~\ref{Fig.1}(a) via a pair of Raman lasers ($\lambda_{3(4)}$) with $\Omega_{3(4)} = 2\pi \times$ 28(21)~MHz) to convert the Rydberg state $|r_1\rangle$ to $|r_2\rangle=|70S_{1/2}\rangle$ via $|e\rangle$ with a detuning $\Delta$ = 2$\pi \times$ 335~MHz, forming $|W_2\rangle$. The two Raman lasers counter-propagate and the $\lambda_3$ field co-propagates with the coupling laser. At the end of the storage of duration $t_s$, we read out $\lvert W_2\rangle$ by applying the retrieval laser $\lambda_5$ with $\Omega_5 = 2\pi \times 16$~MHz.

\begin{figure}[htbp]
    \centering    \includegraphics[width=\linewidth]{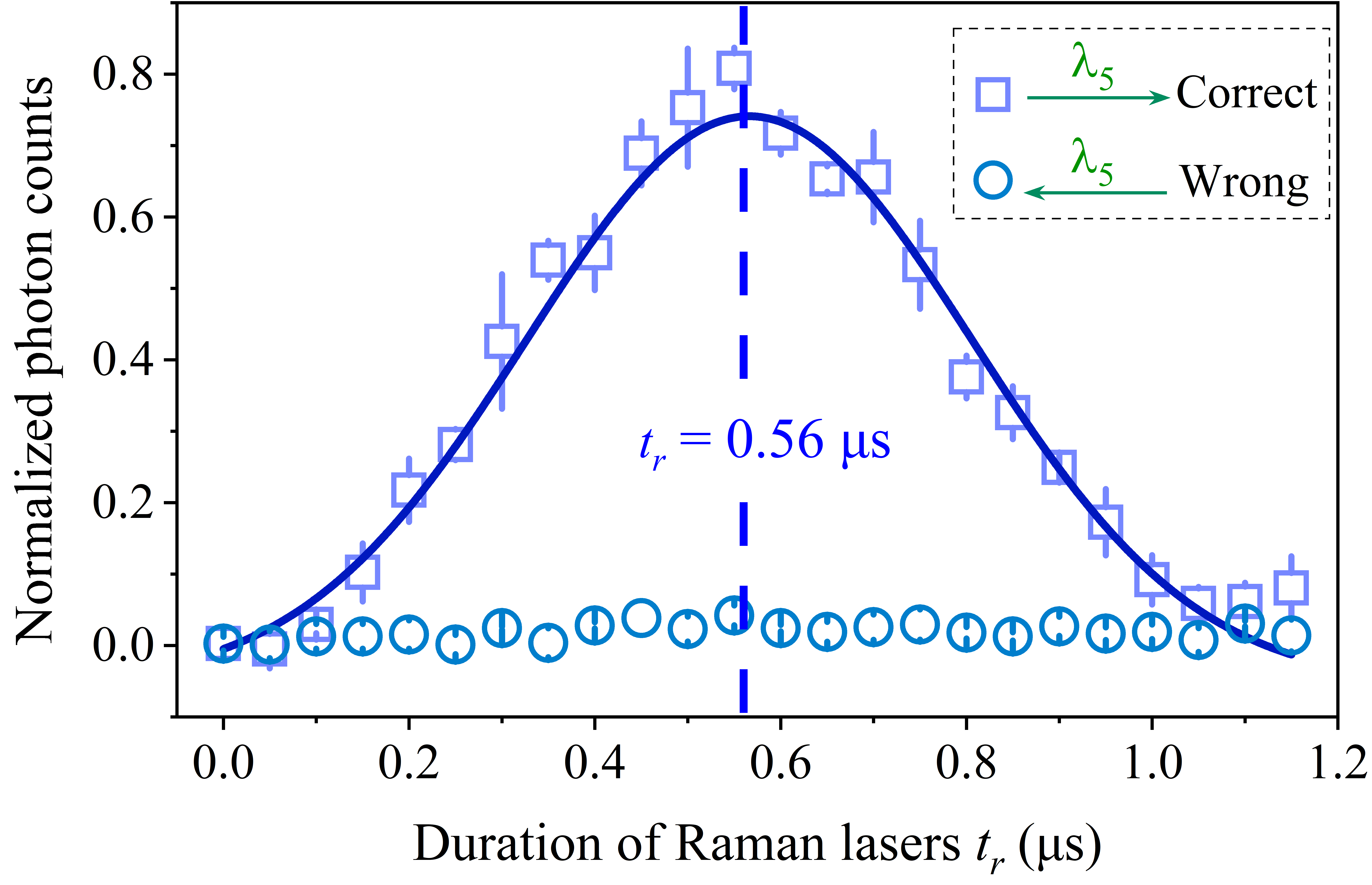}
    \caption{Test of the readout single-photon signal with retrieval laser along 
    different directions. The square (circular) symbols show the retrieval photon counts as a function of the duration of the Raman pulse $t_r$ at $t_s = 7~\mu$s for the correct (wrong) direction of retrieval laser $\lambda_5$, which provides the wavevector matching (mismatching) the triangular relation in Fig.~\ref{Fig.1}. The error bars show the standard deviation of three independent measurements. The fitting curve (solid line) shows that the maximum retrieval efficiency occurs at $t_r = 0.56~\mu$s, corresponding to a $\pi$ pulse.}
    \label{Duration}
\end{figure}

{\it Verification of the feasibility of the protocol --}
Here, we implement a proof-of-principle configuration with case 7 ($\theta_1=0^{\circ}$ and $\theta_2=0^{\circ}$) to demonstrate the feasibility of the protocol. Figure~\ref{Duration} presents the retrieval photon counts as a function of the duration of the Raman laser, $t_r$, for the correct (squares) and wrong direction (circles) with the fixed storage time $t_s = 7~\mu$s. The collective radiation occurs only when the retrieval laser is set in a correct direction as mentioned above, while no photons can be retrieved for the wrong direction of the retrieval laser,  $\theta_2 =\pi$ that the retrieval laser co-propagates with the coupling laser $\lambda_2$.
The solid fitting curve shows that the maximum retrieval efficiency occurs at $t_r = 0.56~\mu$s, which is approximately equal to a $\pi$ pulse duration for the effective Rabi frequency of the Raman laser $\Omega_r = 2\pi \times$ 0.88~MHz. It is worth noting that Fig.~\ref{Duration} demonstrates the readout of photons in the direction along with the direction of the input probe, but, importantly, it verifies the feasibility of the new protocol, which can be easily extended to other cases with suitable laser sources.

\begin{figure}[htbp]
    \centering    \includegraphics[width=\linewidth]{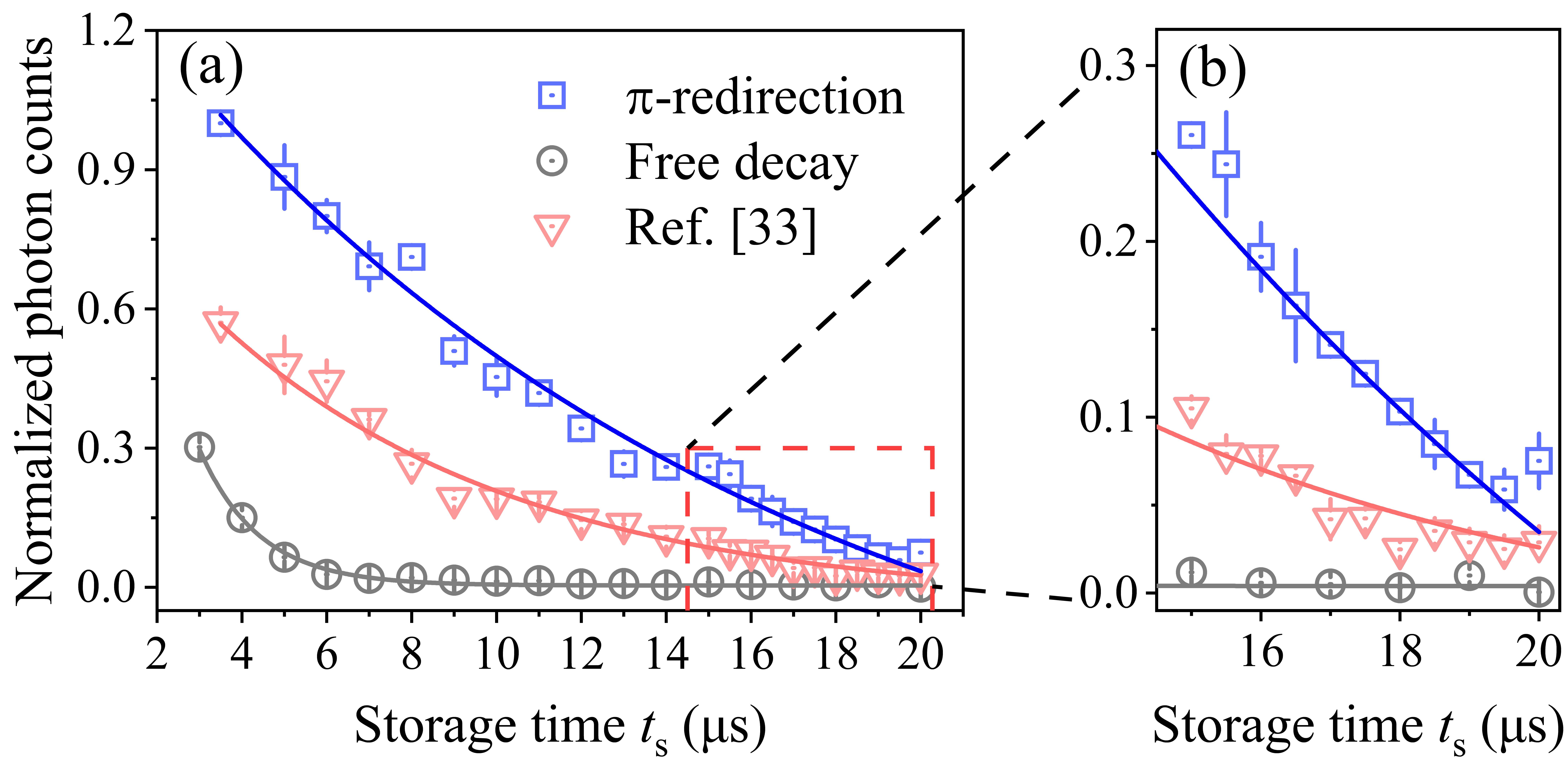}
    \caption{Suppression of motional dephasing by the $\pi$-redirection protocol. (a) Comparison of the retrieval photon counts for the $\pi$-redirection protocol (squares) and the free decay (circles).
    The solid lines show the corresponding fittings. All data are normalized to the first data point of the $\pi$-redirection protocol at $t_s$ = 3.5~$\mu$s. Error bars indicate the standard deviation of three independent measurements. Triangles denote the data of Ref~\cite{PhysRevLett.134.053604} as a comparison.  (b) Zoom-in of the dashed area in Fig.~\ref{Fig.2}(a).}
    \label{Fig.2}
\end{figure}

Moreover, we demonstrate the suppression of the motional dephasing of Rydberg polariton using the $\pi$-redirection protocol.
In Fig.~\ref{Fig.2}(a), we compare the number of retrieval photons for the case of free decay (gray circles) with that of our protocol (blue squares), as a function of storage time $t_s$. The data are normalized to the first data point of the $\pi$-redirection protocol. The optimal wait time $t'$ between the loading of the signal and the start of the $\pi$ pulse is $t_s$-dependent as in Eq.~\ref{protocol3-1}. The retrieval photon counts are almost absent with a storage time beyond 4~$\mu$s for the free decay case, while they are dramatically increased with the $\pi$-redirection protocol. The solid lines correspond to fits with $A_1+B_1\exp{-(t_s/\tau_1)^2}$ and $A_{2}+B_{2}\exp{-(t_s/\tau_{2})}$, yielding $\tau_1=3.11~\mu$s and $\tau_2=14.86~\mu$s, respectively. For comparison, we also present the data using our previous $\pi-$wait$-\pi$ protocol~\cite{PhysRevLett.134.053604}, shown by the red triangles in Fig.~\ref{Fig.2}(a), where all experimental parameters are the same, except for the number difference of $\pi$ pulses. The results clearly show that the use of only one $\pi$ pulse dramatically increases the retrieval efficiency by a factor of 2, as expected theoretically if laser noise is the dominant source of error. In Fig.~\ref{Fig.2}(b), we zoom in on the late time data to demonstrate the effectiveness of the $\pi$-redirection protocol for the long storage times.

{\it Conclusion --} 
In conclusion, we have demonstrated a direction switchable single-photon emitter using Rydberg polaritons. The directional emission is controlled by the direction of the retrieval laser and a suitable choice of the intermediate state of the Raman coupling. Based on this protocol, we propose a generic \textit{N} output single-photon quantum router with unity routing efficiency. 
As the scheme employs a single Raman $\pi$ pulse the effects of laser noise are reduced, resulting in longer coherence times than previous work~\cite{PhysRevLett.134.053604}.

This work was supported by the National Natural Science Foundation of China (No. 12241408, No. 62175136, No. 12120101004, No. U2341211, and No. 12074300); Innovation Program for Quantum Science and Technology (No. 2021ZD0302100, and No. 2023ZD0300902). C.S.A. acknowledges the financial support provided by the UKRI, EPSRC grant Reference No. EP/V030280/1 (“Quantum optics using Rydberg polaritons”). X.-F.S. proposed the protocol with thanks to Yan Lu for fruitful discussions.

\bibliography{main}

\begin{thebibliography}{37}%
\makeatletter
\providecommand \@ifxundefined [1]{%
 \@ifx{#1\undefined}
}%
\providecommand \@ifnum [1]{%
 \ifnum #1\expandafter \@firstoftwo
 \else \expandafter \@secondoftwo
 \fi
}%
\providecommand \@ifx [1]{%
 \ifx #1\expandafter \@firstoftwo
 \else \expandafter \@secondoftwo
 \fi
}%
\providecommand \natexlab [1]{#1}%
\providecommand \enquote  [1]{``#1''}%
\providecommand \bibnamefont  [1]{#1}%
\providecommand \bibfnamefont [1]{#1}%
\providecommand \citenamefont [1]{#1}%
\providecommand \href@noop [0]{\@secondoftwo}%
\providecommand \href [0]{\begingroup \@sanitize@url \@href}%
\providecommand \@href[1]{\@@startlink{#1}\@@href}%
\providecommand \@@href[1]{\endgroup#1\@@endlink}%
\providecommand \@sanitize@url [0]{\catcode `\\12\catcode `\$12\catcode `\&12\catcode `\#12\catcode `\^12\catcode `\_12\catcode `\%12\relax}%
\providecommand \@@startlink[1]{}%
\providecommand \@@endlink[0]{}%
\providecommand \url  [0]{\begingroup\@sanitize@url \@url }%
\providecommand \@url [1]{\endgroup\@href {#1}{\urlprefix }}%
\providecommand \urlprefix  [0]{URL }%
\providecommand \Eprint [0]{\href }%
\providecommand \doibase [0]{https://doi.org/}%
\providecommand \selectlanguage [0]{\@gobble}%
\providecommand \bibinfo  [0]{\@secondoftwo}%
\providecommand \bibfield  [0]{\@secondoftwo}%
\providecommand \translation [1]{[#1]}%
\providecommand \BibitemOpen [0]{}%
\providecommand \bibitemStop [0]{}%
\providecommand \bibitemNoStop [0]{.\EOS\space}%
\providecommand \EOS [0]{\spacefactor3000\relax}%
\providecommand \BibitemShut  [1]{\csname bibitem#1\endcsname}%
\let\auto@bib@innerbib\@empty
\bibitem [{\citenamefont {Monroe}(2002)}]{monroe2002}%
  \BibitemOpen
  \bibfield  {author} {\bibinfo {author} {\bibfnamefont {C.}~\bibnamefont {Monroe}},\ }\bibfield  {title} {\bibinfo {title} {Quantum information processing with atoms and photons},\ }\href {https://doi.org/10.1038/416238a} {\bibfield  {journal} {\bibinfo  {journal} {Nature}\ }\textbf {\bibinfo {volume} {416}},\ \bibinfo {pages} {238} (\bibinfo {year} {2002})}\BibitemShut {NoStop}%
\bibitem [{\citenamefont {Northup}\ and\ \citenamefont {Blatt}(2014)}]{northup2014}%
  \BibitemOpen
  \bibfield  {author} {\bibinfo {author} {\bibfnamefont {T.~E.}\ \bibnamefont {Northup}}\ and\ \bibinfo {author} {\bibfnamefont {R.}~\bibnamefont {Blatt}},\ }\bibfield  {title} {\bibinfo {title} {Quantum information transfer using photons},\ }\href {https://doi.org/10.1038/nphoton.2014.53} {\bibfield  {journal} {\bibinfo  {journal} {Nature Photonics}\ }\textbf {\bibinfo {volume} {8}},\ \bibinfo {pages} {356} (\bibinfo {year} {2014})}\BibitemShut {NoStop}%
\bibitem [{\citenamefont {Kimble}(2008)}]{kimble2008}%
  \BibitemOpen
  \bibfield  {author} {\bibinfo {author} {\bibfnamefont {H.~J.}\ \bibnamefont {Kimble}},\ }\bibfield  {title} {\bibinfo {title} {The quantum internet},\ }\href {https://doi.org/10.1038/nature07127} {\bibfield  {journal} {\bibinfo  {journal} {Nature}\ }\textbf {\bibinfo {volume} {453}},\ \bibinfo {pages} {1023} (\bibinfo {year} {2008})}\BibitemShut {NoStop}%
\bibitem [{\citenamefont {Zhou}\ \emph {et~al.}(2013)\citenamefont {Zhou}, \citenamefont {Yang}, \citenamefont {Li},\ and\ \citenamefont {Sun}}]{PhysRevLett.111.103604}%
  \BibitemOpen
  \bibfield  {author} {\bibinfo {author} {\bibfnamefont {L.}~\bibnamefont {Zhou}}, \bibinfo {author} {\bibfnamefont {L.-P.}\ \bibnamefont {Yang}}, \bibinfo {author} {\bibfnamefont {Y.}~\bibnamefont {Li}},\ and\ \bibinfo {author} {\bibfnamefont {C.~P.}\ \bibnamefont {Sun}},\ }\bibfield  {title} {\bibinfo {title} {Quantum routing of single photons with a cyclic three-level system},\ }\href {https://doi.org/10.1103/PhysRevLett.111.103604} {\bibfield  {journal} {\bibinfo  {journal} {Phys. Rev. Lett.}\ }\textbf {\bibinfo {volume} {111}},\ \bibinfo {pages} {103604} (\bibinfo {year} {2013})}\BibitemShut {NoStop}%
\bibitem [{\citenamefont {Lu}\ \emph {et~al.}(2014)\citenamefont {Lu}, \citenamefont {Zhou}, \citenamefont {Kuang},\ and\ \citenamefont {Nori}}]{PhysRevA.89.013805}%
  \BibitemOpen
  \bibfield  {author} {\bibinfo {author} {\bibfnamefont {J.}~\bibnamefont {Lu}}, \bibinfo {author} {\bibfnamefont {L.}~\bibnamefont {Zhou}}, \bibinfo {author} {\bibfnamefont {L.-M.}\ \bibnamefont {Kuang}},\ and\ \bibinfo {author} {\bibfnamefont {F.}~\bibnamefont {Nori}},\ }\bibfield  {title} {\bibinfo {title} {Single-photon router: Coherent control of multichannel scattering for single photons with quantum interferences},\ }\href {https://doi.org/10.1103/PhysRevA.89.013805} {\bibfield  {journal} {\bibinfo  {journal} {Phys. Rev. A}\ }\textbf {\bibinfo {volume} {89}},\ \bibinfo {pages} {013805} (\bibinfo {year} {2014})}\BibitemShut {NoStop}%
\bibitem [{\citenamefont {Ahumada}\ \emph {et~al.}(2019)\citenamefont {Ahumada}, \citenamefont {Orellana}, \citenamefont {Dom\'{\i}nguez-Adame},\ and\ \citenamefont {Malyshev}}]{PhysRevA.99.033827}%
  \BibitemOpen
  \bibfield  {author} {\bibinfo {author} {\bibfnamefont {M.}~\bibnamefont {Ahumada}}, \bibinfo {author} {\bibfnamefont {P.~A.}\ \bibnamefont {Orellana}}, \bibinfo {author} {\bibfnamefont {F.}~\bibnamefont {Dom\'{\i}nguez-Adame}},\ and\ \bibinfo {author} {\bibfnamefont {A.~V.}\ \bibnamefont {Malyshev}},\ }\bibfield  {title} {\bibinfo {title} {Tunable single-photon quantum router},\ }\href {https://doi.org/10.1103/PhysRevA.99.033827} {\bibfield  {journal} {\bibinfo  {journal} {Phys. Rev. A}\ }\textbf {\bibinfo {volume} {99}},\ \bibinfo {pages} {033827} (\bibinfo {year} {2019})}\BibitemShut {NoStop}%
\bibitem [{\citenamefont {Aoki}\ \emph {et~al.}(2009)\citenamefont {Aoki}, \citenamefont {Parkins}, \citenamefont {Alton}, \citenamefont {Regal}, \citenamefont {Dayan}, \citenamefont {Ostby}, \citenamefont {Vahala},\ and\ \citenamefont {Kimble}}]{PhysRevLett.102.083601}%
  \BibitemOpen
  \bibfield  {author} {\bibinfo {author} {\bibfnamefont {T.}~\bibnamefont {Aoki}}, \bibinfo {author} {\bibfnamefont {A.~S.}\ \bibnamefont {Parkins}}, \bibinfo {author} {\bibfnamefont {D.~J.}\ \bibnamefont {Alton}}, \bibinfo {author} {\bibfnamefont {C.~A.}\ \bibnamefont {Regal}}, \bibinfo {author} {\bibfnamefont {B.}~\bibnamefont {Dayan}}, \bibinfo {author} {\bibfnamefont {E.}~\bibnamefont {Ostby}}, \bibinfo {author} {\bibfnamefont {K.~J.}\ \bibnamefont {Vahala}},\ and\ \bibinfo {author} {\bibfnamefont {H.~J.}\ \bibnamefont {Kimble}},\ }\bibfield  {title} {\bibinfo {title} {Efficient routing of single photons by one atom and a microtoroidal cavity},\ }\href {https://doi.org/10.1103/PhysRevLett.102.083601} {\bibfield  {journal} {\bibinfo  {journal} {Phys. Rev. Lett.}\ }\textbf {\bibinfo {volume} {102}},\ \bibinfo {pages} {083601} (\bibinfo {year} {2009})}\BibitemShut {NoStop}%
\bibitem [{\citenamefont {Xia}\ and\ \citenamefont {Twamley}(2013)}]{PhysRevX.3.031013}%
  \BibitemOpen
  \bibfield  {author} {\bibinfo {author} {\bibfnamefont {K.}~\bibnamefont {Xia}}\ and\ \bibinfo {author} {\bibfnamefont {J.}~\bibnamefont {Twamley}},\ }\bibfield  {title} {\bibinfo {title} {All-optical switching and router via the direct quantum control of coupling between cavity modes},\ }\href {https://doi.org/10.1103/PhysRevX.3.031013} {\bibfield  {journal} {\bibinfo  {journal} {Phys. Rev. X}\ }\textbf {\bibinfo {volume} {3}},\ \bibinfo {pages} {031013} (\bibinfo {year} {2013})}\BibitemShut {NoStop}%
\bibitem [{\citenamefont {Shomroni}\ \emph {et~al.}(2014)\citenamefont {Shomroni}, \citenamefont {Rosenblum}, \citenamefont {Lovsky}, \citenamefont {Bechler}, \citenamefont {Guendelman},\ and\ \citenamefont {Dayan}}]{shomroni2014b}%
  \BibitemOpen
  \bibfield  {author} {\bibinfo {author} {\bibfnamefont {I.}~\bibnamefont {Shomroni}}, \bibinfo {author} {\bibfnamefont {S.}~\bibnamefont {Rosenblum}}, \bibinfo {author} {\bibfnamefont {Y.}~\bibnamefont {Lovsky}}, \bibinfo {author} {\bibfnamefont {O.}~\bibnamefont {Bechler}}, \bibinfo {author} {\bibfnamefont {G.}~\bibnamefont {Guendelman}},\ and\ \bibinfo {author} {\bibfnamefont {B.}~\bibnamefont {Dayan}},\ }\bibfield  {title} {\bibinfo {title} {All-optical routing of single photons by a one-atom switch controlled by a single photon},\ }\href {https://doi.org/10.1126/science.1254699} {\bibfield  {journal} {\bibinfo  {journal} {Science}\ }\textbf {\bibinfo {volume} {345}},\ \bibinfo {pages} {903} (\bibinfo {year} {2014})}\BibitemShut {NoStop}%
\bibitem [{\citenamefont {Yan}\ and\ \citenamefont {Fan}(2014)}]{yan2014single}%
  \BibitemOpen
  \bibfield  {author} {\bibinfo {author} {\bibfnamefont {W.-B.}\ \bibnamefont {Yan}}\ and\ \bibinfo {author} {\bibfnamefont {H.}~\bibnamefont {Fan}},\ }\bibfield  {title} {\bibinfo {title} {Single-photon quantum router with multiple output ports},\ }\href {https://doi.org/https://doi.org/10.1038/srep04820} {\bibfield  {journal} {\bibinfo  {journal} {Scientific reports}\ }\textbf {\bibinfo {volume} {4}},\ \bibinfo {pages} {4820} (\bibinfo {year} {2014})}\BibitemShut {NoStop}%
\bibitem [{\citenamefont {Li}\ and\ \citenamefont {Wei}(2015)}]{PhysRevA.92.063836}%
  \BibitemOpen
  \bibfield  {author} {\bibinfo {author} {\bibfnamefont {X.}~\bibnamefont {Li}}\ and\ \bibinfo {author} {\bibfnamefont {L.~F.}\ \bibnamefont {Wei}},\ }\bibfield  {title} {\bibinfo {title} {Designable single-photon quantum routings with atomic mirrors},\ }\href {https://doi.org/10.1103/PhysRevA.92.063836} {\bibfield  {journal} {\bibinfo  {journal} {Phys. Rev. A}\ }\textbf {\bibinfo {volume} {92}},\ \bibinfo {pages} {063836} (\bibinfo {year} {2015})}\BibitemShut {NoStop}%
\bibitem [{\citenamefont {Zhang}\ \emph {et~al.}(2022)\citenamefont {Zhang}, \citenamefont {Zhu}, \citenamefont {Chen}, \citenamefont {Peng}, \citenamefont {Yin}, \citenamefont {Yang}, \citenamefont {Zhao}, \citenamefont {Lu}, \citenamefont {Chai}, \citenamefont {Xiong} \emph {et~al.}}]{zhang2022controllable}%
  \BibitemOpen
  \bibfield  {author} {\bibinfo {author} {\bibfnamefont {Y.}~\bibnamefont {Zhang}}, \bibinfo {author} {\bibfnamefont {Z.}~\bibnamefont {Zhu}}, \bibinfo {author} {\bibfnamefont {K.}~\bibnamefont {Chen}}, \bibinfo {author} {\bibfnamefont {Z.}~\bibnamefont {Peng}}, \bibinfo {author} {\bibfnamefont {W.}~\bibnamefont {Yin}}, \bibinfo {author} {\bibfnamefont {Y.}~\bibnamefont {Yang}}, \bibinfo {author} {\bibfnamefont {Y.}~\bibnamefont {Zhao}}, \bibinfo {author} {\bibfnamefont {Z.}~\bibnamefont {Lu}}, \bibinfo {author} {\bibfnamefont {Y.}~\bibnamefont {Chai}}, \bibinfo {author} {\bibfnamefont {Z.}~\bibnamefont {Xiong}}, \emph {et~al.},\ }\bibfield  {title} {\bibinfo {title} {Controllable single-photon routing between two waveguides by two giant two-level atoms},\ }\href {https://doi.org/https://doi.org/10.3389/fphy.2022.1054299} {\bibfield  {journal} {\bibinfo  {journal} {Frontiers in Physics}\ }\textbf {\bibinfo {volume} {10}},\ \bibinfo {pages} {1054299} (\bibinfo {year} {2022})}\BibitemShut {NoStop}%
\bibitem [{\citenamefont {Hoi}\ \emph {et~al.}(2011)\citenamefont {Hoi}, \citenamefont {Wilson}, \citenamefont {Johansson}, \citenamefont {Palomaki}, \citenamefont {Peropadre},\ and\ \citenamefont {Delsing}}]{PhysRevLett.107.073601}%
  \BibitemOpen
  \bibfield  {author} {\bibinfo {author} {\bibfnamefont {I.-C.}\ \bibnamefont {Hoi}}, \bibinfo {author} {\bibfnamefont {C.~M.}\ \bibnamefont {Wilson}}, \bibinfo {author} {\bibfnamefont {G.}~\bibnamefont {Johansson}}, \bibinfo {author} {\bibfnamefont {T.}~\bibnamefont {Palomaki}}, \bibinfo {author} {\bibfnamefont {B.}~\bibnamefont {Peropadre}},\ and\ \bibinfo {author} {\bibfnamefont {P.}~\bibnamefont {Delsing}},\ }\bibfield  {title} {\bibinfo {title} {Demonstration of a single-photon router in the microwave regime},\ }\href {https://doi.org/10.1103/PhysRevLett.107.073601} {\bibfield  {journal} {\bibinfo  {journal} {Phys. Rev. Lett.}\ }\textbf {\bibinfo {volume} {107}},\ \bibinfo {pages} {073601} (\bibinfo {year} {2011})}\BibitemShut {NoStop}%
\bibitem [{\citenamefont {Wen}\ \emph {et~al.}(2018)\citenamefont {Wen}, \citenamefont {Kockum}, \citenamefont {Ian}, \citenamefont {Chen}, \citenamefont {Nori},\ and\ \citenamefont {Hoi}}]{PhysRevLett.120.063603}%
  \BibitemOpen
  \bibfield  {author} {\bibinfo {author} {\bibfnamefont {P.~Y.}\ \bibnamefont {Wen}}, \bibinfo {author} {\bibfnamefont {A.~F.}\ \bibnamefont {Kockum}}, \bibinfo {author} {\bibfnamefont {H.}~\bibnamefont {Ian}}, \bibinfo {author} {\bibfnamefont {J.~C.}\ \bibnamefont {Chen}}, \bibinfo {author} {\bibfnamefont {F.}~\bibnamefont {Nori}},\ and\ \bibinfo {author} {\bibfnamefont {I.-C.}\ \bibnamefont {Hoi}},\ }\bibfield  {title} {\bibinfo {title} {Reflective amplification without population inversion from a strongly driven superconducting qubit},\ }\href {https://doi.org/10.1103/PhysRevLett.120.063603} {\bibfield  {journal} {\bibinfo  {journal} {Phys. Rev. Lett.}\ }\textbf {\bibinfo {volume} {120}},\ \bibinfo {pages} {063603} (\bibinfo {year} {2018})}\BibitemShut {NoStop}%
\bibitem [{\citenamefont {Wang}\ \emph {et~al.}(2021)\citenamefont {Wang}, \citenamefont {Wu}, \citenamefont {Bao}, \citenamefont {Li}, \citenamefont {Ma}, \citenamefont {Wang}, \citenamefont {Song}, \citenamefont {Zhang},\ and\ \citenamefont {Duan}}]{PhysRevApplied.15.014049}%
  \BibitemOpen
  \bibfield  {author} {\bibinfo {author} {\bibfnamefont {Z.}~\bibnamefont {Wang}}, \bibinfo {author} {\bibfnamefont {Y.}~\bibnamefont {Wu}}, \bibinfo {author} {\bibfnamefont {Z.}~\bibnamefont {Bao}}, \bibinfo {author} {\bibfnamefont {Y.}~\bibnamefont {Li}}, \bibinfo {author} {\bibfnamefont {C.}~\bibnamefont {Ma}}, \bibinfo {author} {\bibfnamefont {H.}~\bibnamefont {Wang}}, \bibinfo {author} {\bibfnamefont {Y.}~\bibnamefont {Song}}, \bibinfo {author} {\bibfnamefont {H.}~\bibnamefont {Zhang}},\ and\ \bibinfo {author} {\bibfnamefont {L.}~\bibnamefont {Duan}},\ }\bibfield  {title} {\bibinfo {title} {Experimental realization of a deterministic quantum router with superconducting quantum circuits},\ }\href {https://doi.org/10.1103/PhysRevApplied.15.014049} {\bibfield  {journal} {\bibinfo  {journal} {Phys. Rev. Appl.}\ }\textbf {\bibinfo {volume} {15}},\ \bibinfo {pages} {014049} (\bibinfo {year} {2021})}\BibitemShut {NoStop}%
\bibitem [{\citenamefont {Firstenberg}\ \emph {et~al.}(2016)\citenamefont {Firstenberg}, \citenamefont {Adams},\ and\ \citenamefont {Hofferberth}}]{Firstenberg_2016}%
  \BibitemOpen
  \bibfield  {author} {\bibinfo {author} {\bibfnamefont {O.}~\bibnamefont {Firstenberg}}, \bibinfo {author} {\bibfnamefont {C.~S.}\ \bibnamefont {Adams}},\ and\ \bibinfo {author} {\bibfnamefont {S.}~\bibnamefont {Hofferberth}},\ }\bibfield  {title} {\bibinfo {title} {Nonlinear quantum optics mediated by rydberg interactions},\ }\href {https://doi.org/10.1088/0953-4075/49/15/152003} {\bibfield  {journal} {\bibinfo  {journal} {Journal of Physics B: Atomic, Molecular and Optical Physics}\ }\textbf {\bibinfo {volume} {49}},\ \bibinfo {pages} {152003} (\bibinfo {year} {2016})}\BibitemShut {NoStop}%
\bibitem [{\citenamefont {Dudin}\ and\ \citenamefont {Kuzmich}(2012)}]{dudin2012Strongly}%
  \BibitemOpen
  \bibfield  {author} {\bibinfo {author} {\bibfnamefont {Y.~O.}\ \bibnamefont {Dudin}}\ and\ \bibinfo {author} {\bibfnamefont {A.}~\bibnamefont {Kuzmich}},\ }\bibfield  {title} {\bibinfo {title} {Strongly {{Interacting Rydberg Excitations}} of a {{Cold Atomic Gas}}},\ }\href {https://doi.org/10.1126/science.1217901} {\bibfield  {journal} {\bibinfo  {journal} {Science}\ }\textbf {\bibinfo {volume} {336}},\ \bibinfo {pages} {887} (\bibinfo {year} {2012})}\BibitemShut {NoStop}%
\bibitem [{\citenamefont {Peyronel}\ \emph {et~al.}(2012)\citenamefont {Peyronel}, \citenamefont {Firstenberg}, \citenamefont {Liang}, \citenamefont {Hofferberth}, \citenamefont {Gorshkov}, \citenamefont {Pohl}, \citenamefont {Lukin},\ and\ \citenamefont {Vuleti{\'c}}}]{peyronel2012Quantumb}%
  \BibitemOpen
  \bibfield  {author} {\bibinfo {author} {\bibfnamefont {T.}~\bibnamefont {Peyronel}}, \bibinfo {author} {\bibfnamefont {O.}~\bibnamefont {Firstenberg}}, \bibinfo {author} {\bibfnamefont {Q.-Y.}\ \bibnamefont {Liang}}, \bibinfo {author} {\bibfnamefont {S.}~\bibnamefont {Hofferberth}}, \bibinfo {author} {\bibfnamefont {A.~V.}\ \bibnamefont {Gorshkov}}, \bibinfo {author} {\bibfnamefont {T.}~\bibnamefont {Pohl}}, \bibinfo {author} {\bibfnamefont {M.~D.}\ \bibnamefont {Lukin}},\ and\ \bibinfo {author} {\bibfnamefont {V.}~\bibnamefont {Vuleti{\'c}}},\ }\bibfield  {title} {\bibinfo {title} {Quantum nonlinear optics with single photons enabled by strongly interacting atoms},\ }\href {https://doi.org/10.1038/nature11361} {\bibfield  {journal} {\bibinfo  {journal} {Nature}\ }\textbf {\bibinfo {volume} {488}},\ \bibinfo {pages} {57} (\bibinfo {year} {2012})}\BibitemShut {NoStop}%
\bibitem [{\citenamefont {Maxwell}\ \emph {et~al.}(2013)\citenamefont {Maxwell}, \citenamefont {Szwer}, \citenamefont {Paredes-Barato}, \citenamefont {Busche}, \citenamefont {Pritchard}, \citenamefont {Gauguet}, \citenamefont {Weatherill}, \citenamefont {Jones},\ and\ \citenamefont {Adams}}]{Maxwell2013}%
  \BibitemOpen
  \bibfield  {author} {\bibinfo {author} {\bibfnamefont {D.}~\bibnamefont {Maxwell}}, \bibinfo {author} {\bibfnamefont {D.~J.}\ \bibnamefont {Szwer}}, \bibinfo {author} {\bibfnamefont {D.}~\bibnamefont {Paredes-Barato}}, \bibinfo {author} {\bibfnamefont {H.}~\bibnamefont {Busche}}, \bibinfo {author} {\bibfnamefont {J.~D.}\ \bibnamefont {Pritchard}}, \bibinfo {author} {\bibfnamefont {A.}~\bibnamefont {Gauguet}}, \bibinfo {author} {\bibfnamefont {K.~J.}\ \bibnamefont {Weatherill}}, \bibinfo {author} {\bibfnamefont {M.~P.~A.}\ \bibnamefont {Jones}},\ and\ \bibinfo {author} {\bibfnamefont {C.~S.}\ \bibnamefont {Adams}},\ }\bibfield  {title} {\bibinfo {title} {Storage and control of optical photons using rydberg polaritons},\ }\href {https://doi.org/10.1103/PhysRevLett.110.103001} {\bibfield  {journal} {\bibinfo  {journal} {Phys. Rev. Lett.}\ }\textbf {\bibinfo {volume} {110}},\ \bibinfo {pages} {103001} (\bibinfo {year} {2013})}\BibitemShut {NoStop}%
\bibitem [{\citenamefont {{Ornelas-Huerta}}\ \emph {et~al.}(2020)\citenamefont {{Ornelas-Huerta}}, \citenamefont {Craddock}, \citenamefont {Goldschmidt}, \citenamefont {Hachtel}, \citenamefont {Wang}, \citenamefont {Bienias}, \citenamefont {Gorshkov}, \citenamefont {Rolston},\ and\ \citenamefont {Porto}}]{ornelas-huerta2020Ondemand}%
  \BibitemOpen
  \bibfield  {author} {\bibinfo {author} {\bibfnamefont {D.~P.}\ \bibnamefont {{Ornelas-Huerta}}}, \bibinfo {author} {\bibfnamefont {A.~N.}\ \bibnamefont {Craddock}}, \bibinfo {author} {\bibfnamefont {E.~A.}\ \bibnamefont {Goldschmidt}}, \bibinfo {author} {\bibfnamefont {A.~J.}\ \bibnamefont {Hachtel}}, \bibinfo {author} {\bibfnamefont {Y.}~\bibnamefont {Wang}}, \bibinfo {author} {\bibfnamefont {P.}~\bibnamefont {Bienias}}, \bibinfo {author} {\bibfnamefont {A.~V.}\ \bibnamefont {Gorshkov}}, \bibinfo {author} {\bibfnamefont {S.~L.}\ \bibnamefont {Rolston}},\ and\ \bibinfo {author} {\bibfnamefont {J.~V.}\ \bibnamefont {Porto}},\ }\bibfield  {title} {\bibinfo {title} {On-demand indistinguishable single photons from an efficient and pure source based on a {{Rydberg}} ensemble},\ }\href {https://doi.org/10.1364/OPTICA.391485} {\bibfield  {journal} {\bibinfo  {journal} {Optica}\ }\textbf {\bibinfo {volume} {7}},\ \bibinfo {pages} {813} (\bibinfo {year} {2020})}\BibitemShut {NoStop}%
\bibitem [{\citenamefont {{Padr{\'o}n-Brito}}\ \emph {et~al.}(2021)\citenamefont {{Padr{\'o}n-Brito}}, \citenamefont {Lowinski}, \citenamefont {Farrera}, \citenamefont {Theophilo},\ and\ \citenamefont {{de Riedmatten}}}]{padron-brito2021Probing}%
  \BibitemOpen
  \bibfield  {author} {\bibinfo {author} {\bibfnamefont {A.}~\bibnamefont {{Padr{\'o}n-Brito}}}, \bibinfo {author} {\bibfnamefont {J.}~\bibnamefont {Lowinski}}, \bibinfo {author} {\bibfnamefont {P.}~\bibnamefont {Farrera}}, \bibinfo {author} {\bibfnamefont {K.}~\bibnamefont {Theophilo}},\ and\ \bibinfo {author} {\bibfnamefont {H.}~\bibnamefont {{de Riedmatten}}},\ }\bibfield  {title} {\bibinfo {title} {Probing the indistinguishability of single photons generated by {{Rydberg}} atomic ensembles},\ }\href {https://doi.org/10.1103/PhysRevResearch.3.033287} {\bibfield  {journal} {\bibinfo  {journal} {Phys. Rev. Research}\ }\textbf {\bibinfo {volume} {3}},\ \bibinfo {pages} {033287} (\bibinfo {year} {2021})}\BibitemShut {NoStop}%
\bibitem [{\citenamefont {Baur}\ \emph {et~al.}(2014)\citenamefont {Baur}, \citenamefont {Tiarks}, \citenamefont {Rempe},\ and\ \citenamefont {D{\"u}rr}}]{baur2014SinglePhotonb}%
  \BibitemOpen
  \bibfield  {author} {\bibinfo {author} {\bibfnamefont {S.}~\bibnamefont {Baur}}, \bibinfo {author} {\bibfnamefont {D.}~\bibnamefont {Tiarks}}, \bibinfo {author} {\bibfnamefont {G.}~\bibnamefont {Rempe}},\ and\ \bibinfo {author} {\bibfnamefont {S.}~\bibnamefont {D{\"u}rr}},\ }\bibfield  {title} {\bibinfo {title} {Single-{{Photon Switch Based}} on {{Rydberg Blockade}}},\ }\href {https://doi.org/10.1103/PhysRevLett.112.073901} {\bibfield  {journal} {\bibinfo  {journal} {Phys. Rev. Lett.}\ }\textbf {\bibinfo {volume} {112}},\ \bibinfo {pages} {073901} (\bibinfo {year} {2014})}\BibitemShut {NoStop}%
\bibitem [{\citenamefont {Tiarks}\ \emph {et~al.}(2014)\citenamefont {Tiarks}, \citenamefont {Baur}, \citenamefont {Schneider}, \citenamefont {D{\"u}rr},\ and\ \citenamefont {Rempe}}]{tiarks2014SinglePhoton}%
  \BibitemOpen
  \bibfield  {author} {\bibinfo {author} {\bibfnamefont {D.}~\bibnamefont {Tiarks}}, \bibinfo {author} {\bibfnamefont {S.}~\bibnamefont {Baur}}, \bibinfo {author} {\bibfnamefont {K.}~\bibnamefont {Schneider}}, \bibinfo {author} {\bibfnamefont {S.}~\bibnamefont {D{\"u}rr}},\ and\ \bibinfo {author} {\bibfnamefont {G.}~\bibnamefont {Rempe}},\ }\bibfield  {title} {\bibinfo {title} {Single-{{Photon Transistor Using}} a {{F{\"o}rster Resonance}}},\ }\href {https://doi.org/10.1103/PhysRevLett.113.053602} {\bibfield  {journal} {\bibinfo  {journal} {Phys. Rev. Lett.}\ }\textbf {\bibinfo {volume} {113}},\ \bibinfo {pages} {053602} (\bibinfo {year} {2014})}\BibitemShut {NoStop}%
\bibitem [{\citenamefont {Gorniaczyk}\ \emph {et~al.}(2014)\citenamefont {Gorniaczyk}, \citenamefont {Tresp}, \citenamefont {Schmidt}, \citenamefont {Fedder},\ and\ \citenamefont {Hofferberth}}]{gorniaczyk2014SinglePhoton}%
  \BibitemOpen
  \bibfield  {author} {\bibinfo {author} {\bibfnamefont {H.}~\bibnamefont {Gorniaczyk}}, \bibinfo {author} {\bibfnamefont {C.}~\bibnamefont {Tresp}}, \bibinfo {author} {\bibfnamefont {J.}~\bibnamefont {Schmidt}}, \bibinfo {author} {\bibfnamefont {H.}~\bibnamefont {Fedder}},\ and\ \bibinfo {author} {\bibfnamefont {S.}~\bibnamefont {Hofferberth}},\ }\bibfield  {title} {\bibinfo {title} {Single-{{Photon Transistor Mediated}} by {{Interstate Rydberg Interactions}}},\ }\href {https://doi.org/10.1103/PhysRevLett.113.053601} {\bibfield  {journal} {\bibinfo  {journal} {Phys. Rev. Lett.}\ }\textbf {\bibinfo {volume} {113}},\ \bibinfo {pages} {053601} (\bibinfo {year} {2014})}\BibitemShut {NoStop}%
\bibitem [{\citenamefont {Tiarks}\ \emph {et~al.}(2019)\citenamefont {Tiarks}, \citenamefont {{Schmidt-Eberle}}, \citenamefont {Stolz}, \citenamefont {Rempe},\ and\ \citenamefont {D{\"u}rr}}]{tiarks2019Photon}%
  \BibitemOpen
  \bibfield  {author} {\bibinfo {author} {\bibfnamefont {D.}~\bibnamefont {Tiarks}}, \bibinfo {author} {\bibfnamefont {S.}~\bibnamefont {{Schmidt-Eberle}}}, \bibinfo {author} {\bibfnamefont {T.}~\bibnamefont {Stolz}}, \bibinfo {author} {\bibfnamefont {G.}~\bibnamefont {Rempe}},\ and\ \bibinfo {author} {\bibfnamefont {S.}~\bibnamefont {D{\"u}rr}},\ }\bibfield  {title} {\bibinfo {title} {A photon--photon quantum gate based on {{Rydberg}} interactions},\ }\href {https://doi.org/10.1038/s41567-018-0313-7} {\bibfield  {journal} {\bibinfo  {journal} {Nat. Phys.}\ }\textbf {\bibinfo {volume} {15}},\ \bibinfo {pages} {124} (\bibinfo {year} {2019})}\BibitemShut {NoStop}%
\bibitem [{\citenamefont {Shi}\ \emph {et~al.}(2022)\citenamefont {Shi}, \citenamefont {Xu}, \citenamefont {Zhang}, \citenamefont {Ye}, \citenamefont {Xiang}, \citenamefont {Liu}, \citenamefont {Wang}, \citenamefont {Su},\ and\ \citenamefont {Li}}]{shi2022Highfidelityb}%
  \BibitemOpen
  \bibfield  {author} {\bibinfo {author} {\bibfnamefont {S.}~\bibnamefont {Shi}}, \bibinfo {author} {\bibfnamefont {B.}~\bibnamefont {Xu}}, \bibinfo {author} {\bibfnamefont {K.}~\bibnamefont {Zhang}}, \bibinfo {author} {\bibfnamefont {G.-S.}\ \bibnamefont {Ye}}, \bibinfo {author} {\bibfnamefont {D.-S.}\ \bibnamefont {Xiang}}, \bibinfo {author} {\bibfnamefont {Y.}~\bibnamefont {Liu}}, \bibinfo {author} {\bibfnamefont {J.}~\bibnamefont {Wang}}, \bibinfo {author} {\bibfnamefont {D.}~\bibnamefont {Su}},\ and\ \bibinfo {author} {\bibfnamefont {L.}~\bibnamefont {Li}},\ }\bibfield  {title} {\bibinfo {title} {High-fidelity photonic quantum logic gate based on near-optimal {{Rydberg}} single-photon source},\ }\href {https://doi.org/10.1038/s41467-022-32083-9} {\bibfield  {journal} {\bibinfo  {journal} {Nat. Commun.}\ }\textbf {\bibinfo {volume} {13}},\ \bibinfo {pages} {4454} (\bibinfo {year} {2022})}\BibitemShut {NoStop}%
\bibitem [{\citenamefont {Ye}\ \emph {et~al.}(2023)\citenamefont {Ye}, \citenamefont {Xu}, \citenamefont {Chang}, \citenamefont {Shi}, \citenamefont {Shi},\ and\ \citenamefont {Li}}]{ye2023Photonica}%
  \BibitemOpen
  \bibfield  {author} {\bibinfo {author} {\bibfnamefont {G.-S.}\ \bibnamefont {Ye}}, \bibinfo {author} {\bibfnamefont {B.}~\bibnamefont {Xu}}, \bibinfo {author} {\bibfnamefont {Y.}~\bibnamefont {Chang}}, \bibinfo {author} {\bibfnamefont {S.}~\bibnamefont {Shi}}, \bibinfo {author} {\bibfnamefont {T.}~\bibnamefont {Shi}},\ and\ \bibinfo {author} {\bibfnamefont {L.}~\bibnamefont {Li}},\ }\bibfield  {title} {\bibinfo {title} {A photonic entanglement filter with {{Rydberg}} atoms},\ }\href {https://doi.org/10.1038/s41566-023-01194-0} {\bibfield  {journal} {\bibinfo  {journal} {Nat. Photonics}\ }\textbf {\bibinfo {volume} {17}},\ \bibinfo {pages} {538} (\bibinfo {year} {2023})}\BibitemShut {NoStop}%
\bibitem [{\citenamefont {Busche}\ \emph {et~al.}(2017)\citenamefont {Busche}, \citenamefont {Huillery}, \citenamefont {Ball}, \citenamefont {Ilieva}, \citenamefont {Jones},\ and\ \citenamefont {Adams}}]{busche2017Contactlessb}%
  \BibitemOpen
  \bibfield  {author} {\bibinfo {author} {\bibfnamefont {H.}~\bibnamefont {Busche}}, \bibinfo {author} {\bibfnamefont {P.}~\bibnamefont {Huillery}}, \bibinfo {author} {\bibfnamefont {S.~W.}\ \bibnamefont {Ball}}, \bibinfo {author} {\bibfnamefont {T.}~\bibnamefont {Ilieva}}, \bibinfo {author} {\bibfnamefont {M.~P.~A.}\ \bibnamefont {Jones}},\ and\ \bibinfo {author} {\bibfnamefont {C.~S.}\ \bibnamefont {Adams}},\ }\bibfield  {title} {\bibinfo {title} {Contactless nonlinear optics mediated by long-range {{Rydberg}} interactions},\ }\href {https://doi.org/10.1038/nphys4058} {\bibfield  {journal} {\bibinfo  {journal} {Nat. Phys.}\ }\textbf {\bibinfo {volume} {13}},\ \bibinfo {pages} {655} (\bibinfo {year} {2017})}\BibitemShut {NoStop}%
\bibitem [{\citenamefont {Spong}\ \emph {et~al.}(2021)\citenamefont {Spong}, \citenamefont {Jiao}, \citenamefont {Hughes}, \citenamefont {Weatherill}, \citenamefont {Lesanovsky},\ and\ \citenamefont {Adams}}]{spong2021collectively}%
  \BibitemOpen
  \bibfield  {author} {\bibinfo {author} {\bibfnamefont {N.~L.~R.}\ \bibnamefont {Spong}}, \bibinfo {author} {\bibfnamefont {Y.}~\bibnamefont {Jiao}}, \bibinfo {author} {\bibfnamefont {O.~D.~W.}\ \bibnamefont {Hughes}}, \bibinfo {author} {\bibfnamefont {K.~J.}\ \bibnamefont {Weatherill}}, \bibinfo {author} {\bibfnamefont {I.}~\bibnamefont {Lesanovsky}},\ and\ \bibinfo {author} {\bibfnamefont {C.~S.}\ \bibnamefont {Adams}},\ }\bibfield  {title} {\bibinfo {title} {Collectively {{Encoded Rydberg Qubit}}},\ }\href {https://doi.org/10.1103/PhysRevLett.127.063604} {\bibfield  {journal} {\bibinfo  {journal} {Phys. Rev. Lett.}\ }\textbf {\bibinfo {volume} {127}},\ \bibinfo {pages} {063604} (\bibinfo {year} {2021})}\BibitemShut {NoStop}%
\bibitem [{\citenamefont {Murray}\ and\ \citenamefont {Pohl}(2017)}]{murray2017a}%
  \BibitemOpen
  \bibfield  {author} {\bibinfo {author} {\bibfnamefont {C.~R.}\ \bibnamefont {Murray}}\ and\ \bibinfo {author} {\bibfnamefont {T.}~\bibnamefont {Pohl}},\ }\bibfield  {title} {\bibinfo {title} {Coherent {{Photon Manipulation}} in {{Interacting Atomic Ensembles}}},\ }\href {https://doi.org/10.1103/PhysRevX.7.031007} {\bibfield  {journal} {\bibinfo  {journal} {Physical Review X}\ }\textbf {\bibinfo {volume} {7}},\ \bibinfo {pages} {031007} (\bibinfo {year} {2017})}\BibitemShut {NoStop}%
\bibitem [{\citenamefont {Ou}\ \emph {et~al.}(2022)\citenamefont {Ou}, \citenamefont {Zhang},\ and\ \citenamefont {Huang}}]{ou2022}%
  \BibitemOpen
  \bibfield  {author} {\bibinfo {author} {\bibfnamefont {Y.}~\bibnamefont {Ou}}, \bibinfo {author} {\bibfnamefont {Q.}~\bibnamefont {Zhang}},\ and\ \bibinfo {author} {\bibfnamefont {G.}~\bibnamefont {Huang}},\ }\bibfield  {title} {\bibinfo {title} {Quantum reflection of single photons in a cold {{Rydberg}} atomic gas},\ }\href {https://doi.org/10.1364/OL.469347} {\bibfield  {journal} {\bibinfo  {journal} {Optics Letters}\ }\textbf {\bibinfo {volume} {47}},\ \bibinfo {pages} {4395} (\bibinfo {year} {2022})}\BibitemShut {NoStop}%
\bibitem [{\citenamefont {Palaiodimopoulos}\ \emph {et~al.}(2024)\citenamefont {Palaiodimopoulos}, \citenamefont {Ohler}, \citenamefont {Fleischhauer},\ and\ \citenamefont {Petrosyan}}]{palaiodimopoulos2024}%
  \BibitemOpen
  \bibfield  {author} {\bibinfo {author} {\bibfnamefont {N.~E.}\ \bibnamefont {Palaiodimopoulos}}, \bibinfo {author} {\bibfnamefont {S.}~\bibnamefont {Ohler}}, \bibinfo {author} {\bibfnamefont {M.}~\bibnamefont {Fleischhauer}},\ and\ \bibinfo {author} {\bibfnamefont {D.}~\bibnamefont {Petrosyan}},\ }\bibfield  {title} {\bibinfo {title} {Chiral quantum router with {{Rydberg}} atoms},\ }\href {https://doi.org/10.1103/PhysRevA.109.032622} {\bibfield  {journal} {\bibinfo  {journal} {Physical Review A}\ }\textbf {\bibinfo {volume} {109}},\ \bibinfo {pages} {032622} (\bibinfo {year} {2024})}\BibitemShut {NoStop}%
\bibitem [{\citenamefont {Du}\ \emph {et~al.}(2021)\citenamefont {Du}, \citenamefont {Cao}, \citenamefont {Dang},\ and\ \citenamefont {Tan}}]{du2021}%
  \BibitemOpen
  \bibfield  {author} {\bibinfo {author} {\bibfnamefont {X.-P.}\ \bibnamefont {Du}}, \bibinfo {author} {\bibfnamefont {Q.}~\bibnamefont {Cao}}, \bibinfo {author} {\bibfnamefont {N.}~\bibnamefont {Dang}},\ and\ \bibinfo {author} {\bibfnamefont {L.}~\bibnamefont {Tan}},\ }\bibfield  {title} {\bibinfo {title} {Quantum router modulated by two {{Rydberg}} atoms in a {{X-shaped}} coupled cavity array},\ }\href {https://doi.org/10.1140/epjd/s10053-021-00085-9} {\bibfield  {journal} {\bibinfo  {journal} {The European Physical Journal D}\ }\textbf {\bibinfo {volume} {75}},\ \bibinfo {pages} {79} (\bibinfo {year} {2021})}\BibitemShut {NoStop}%
\bibitem [{\citenamefont {Jiao}\ \emph {et~al.}(2025)\citenamefont {Jiao}, \citenamefont {Li}, \citenamefont {Shi}, \citenamefont {Fan}, \citenamefont {Bai}, \citenamefont {Jia}, \citenamefont {Zhao},\ and\ \citenamefont {Adams}}]{PhysRevLett.134.053604}%
  \BibitemOpen
  \bibfield  {author} {\bibinfo {author} {\bibfnamefont {Y.}~\bibnamefont {Jiao}}, \bibinfo {author} {\bibfnamefont {C.}~\bibnamefont {Li}}, \bibinfo {author} {\bibfnamefont {X.-F.}\ \bibnamefont {Shi}}, \bibinfo {author} {\bibfnamefont {J.}~\bibnamefont {Fan}}, \bibinfo {author} {\bibfnamefont {J.}~\bibnamefont {Bai}}, \bibinfo {author} {\bibfnamefont {S.}~\bibnamefont {Jia}}, \bibinfo {author} {\bibfnamefont {J.}~\bibnamefont {Zhao}},\ and\ \bibinfo {author} {\bibfnamefont {C.~S.}\ \bibnamefont {Adams}},\ }\bibfield  {title} {\bibinfo {title} {Suppression of motional dephasing using state mapping},\ }\href {https://doi.org/10.1103/PhysRevLett.134.053604} {\bibfield  {journal} {\bibinfo  {journal} {Phys. Rev. Lett.}\ }\textbf {\bibinfo {volume} {134}},\ \bibinfo {pages} {053604} (\bibinfo {year} {2025})}\BibitemShut {NoStop}%
\bibitem [{\citenamefont {Fleischhauer}\ and\ \citenamefont {Lukin}(2002)}]{Fleischhauer2002}%
  \BibitemOpen
  \bibfield  {author} {\bibinfo {author} {\bibfnamefont {M.}~\bibnamefont {Fleischhauer}}\ and\ \bibinfo {author} {\bibfnamefont {M.~D.}\ \bibnamefont {Lukin}},\ }\bibfield  {title} {\bibinfo {title} {{Quantum memory for photons: Dark-state polaritons}},\ }\href {https://doi.org/10.1103/PhysRevA.65.022314} {\bibfield  {journal} {\bibinfo  {journal} {Phys. Rev. A}\ }\textbf {\bibinfo {volume} {65}},\ \bibinfo {pages} {022314} (\bibinfo {year} {2002})}\BibitemShut {NoStop}%
\bibitem [{\citenamefont {Li}\ and\ \citenamefont {Kuzmich}(2016)}]{li2016Quantuma}%
  \BibitemOpen
  \bibfield  {author} {\bibinfo {author} {\bibfnamefont {L.}~\bibnamefont {Li}}\ and\ \bibinfo {author} {\bibfnamefont {A.}~\bibnamefont {Kuzmich}},\ }\bibfield  {title} {\bibinfo {title} {Quantum memory with strong and controllable {{Rydberg-level}} interactions},\ }\href {https://doi.org/10.1038/ncomms13618} {\bibfield  {journal} {\bibinfo  {journal} {Nat. Commun.}\ }\textbf {\bibinfo {volume} {7}},\ \bibinfo {pages} {13618} (\bibinfo {year} {2016})}\BibitemShut {NoStop}%
\bibitem [{\citenamefont {Adams}\ \emph {et~al.}(2020)\citenamefont {Adams}, \citenamefont {Pritchard},\ and\ \citenamefont {Shaffer}}]{Adams2020}%
  \BibitemOpen
  \bibfield  {author} {\bibinfo {author} {\bibfnamefont {C.~S.}\ \bibnamefont {Adams}}, \bibinfo {author} {\bibfnamefont {J.~D.}\ \bibnamefont {Pritchard}},\ and\ \bibinfo {author} {\bibfnamefont {J.~P.}\ \bibnamefont {Shaffer}},\ }\bibfield  {title} {\bibinfo {title} {Rydberg atom quantum technologies},\ }\href {https://doi.org/10.1088/1361-6455/ab52ef} {\bibfield  {journal} {\bibinfo  {journal} {J. Phys. B: At. Mol. Opt. Phys.}\ }\textbf {\bibinfo {volume} {53}},\ \bibinfo {pages} {012002} (\bibinfo {year} {2020})}\BibitemShut {NoStop}%
\end{thebibliography}%

\end{document}